\documentclass[letter,longauth]{aa}
\usepackage{graphicx}
\usepackage{txfonts}

\usepackage{bm}

\usepackage[]{hyperref}
\begin{document}

   \title{First direct detection of an exoplanet by optical interferometry\thanks{Based on observations collected at the European Organisation for Astronomical Research in the Southern Hemisphere, ID 60.A-9102(G).
   }}
     \subtitle{Astrometry and K-band spectroscopy of HR8799\,e}
\titlerunning{Astrometry and K-band spectroscopy of HR8799\,e}

\author{GRAVITY Collaboration\thanks{ Corresponding authors \email{sylvestre.lacour@obspm.fr}, and \email{mathias.nowak@obspm.fr}. 
}:
   S. Lacour\inst{1,2}
       \and
   M. Nowak
             \inst{1}         \and
   J. Wang
             \inst{20,}\thanks{51 Pegasi b Fellow.}         \and
  O.~Pfuhl
             \inst{2}       \and
  F.~Eisenhauer
             \inst{2}
\and R.~Abuter\inst{8}
\and A.~Amorim\inst{6,14}
\and N.~Anugu\inst{7,22}
\and M.~Benisty\inst{5}
\and J.P.~Berger\inst{5}
\and H.~Beust  \inst{5}
\and N.~Blind\inst{10}
\and M.~Bonnefoy   \inst{5}
\and H.~Bonnet\inst{8}
\and P.~Bourget\inst{9}
\and W.~Brandner\inst{3}
\and A.~Buron\inst{2}
\and C.~Collin\inst{1}
\and B.~Charnay\inst{1}
\and F.~Chapron\inst{1}
\and Y.~Cl\'{e}net\inst{1}
\and V.~Coud\'e~du~Foresto\inst{1}
\and P.T.~de~Zeeuw\inst{2,12}
\and C.~Deen\inst{2}
\and R.~Dembet\inst{1}
\and J.~Dexter\inst{2}
\and G.~Duvert\inst{5}
\and A.~Eckart\inst{4,11}
\and N.M.~Förster~Schreiber\inst{2}
\and P.~Fédou\inst{1}
\and P.~Garcia\inst{7,9,14}
\and R.~Garcia~Lopez\inst{15,3}
\and F.~Gao\inst{2}
\and E.~Gendron\inst{1}
\and R.~Genzel\inst{2,13}
\and S.~Gillessen\inst{2}
\and P.~Gordo\inst{6,14}
\and  A.~Greenbaum \inst{16}   
\and M.~Habibi\inst{2}
\and X.~Haubois\inst{9}
\and F.~Haußmann\inst{2}
\and Th.~Henning\inst{3}
\and S.~Hippler\inst{3}
\and M.~Horrobin\inst{4}
\and Z.~Hubert\inst{1}
\and A.~Jimenez~Rosales\inst{2}
\and L.~Jocou\inst{5}
\and S.~Kendrew\inst{17,3}
\and P.~Kervella\inst{1}
\and J.~Kolb\inst{9} 
\and  A.-M.~Lagrange   \inst{5}
\and V.~Lapeyr\`ere\inst{1}
\and J.-B.~Le~Bouquin\inst{5}
\and P.~L\'ena\inst{1}
\and M.~Lippa\inst{2}
\and R.~Lenzen\inst{3}
\and A.-L.~Maire\inst{19,3}
\and  P.~Mollière   \inst{12}    
\and T.~Ott\inst{2}
\and T.~Paumard\inst{1}
\and K.~Perraut\inst{5}
\and G.~Perrin\inst{1}
\and  L.~Pueyo  \inst{18}  
\and S.~Rabien\inst{2}
\and A.~Ram\'irez\inst{9} 
\and C.~Rau\inst{2}
\and G.~Rodr\'iguez-Coira\inst{1}
\and G.~Rousset\inst{1}
\and J.~Sanchez-Bermudez\inst{21,3}
\and S.~Scheithauer\inst{3}
\and N.~Schuhler\inst{9} 
\and O.~Straub\inst{1,2}
\and C.~Straubmeier\inst{4}
\and E.~Sturm\inst{2}
\and L.J.~Tacconi\inst{2}
\and F.~Vincent\inst{1}
\and E.F.~van~Dishoeck   \inst{2,12}
\and S.~von~Fellenberg\inst{2}
\and I.~Wank\inst{4}
\and I.~Waisberg\inst{2}
\and F.~Widmann\inst{2}
\and E.~Wieprecht\inst{2}
\and M.~Wiest\inst{4}
\and E.~Wiezorrek\inst{2}
\and J.~Woillez\inst{8}
\and S.~Yazici\inst{2,4}
\and D.~Ziegler\inst{1}
\and G.~Zins\inst{9}
          }

   \institute{
    LESIA, Observatoire de Paris, Universit\'e PSL,
CNRS, Sorbonne Universit\'e, Univ. Paris Diderot,
Sorbonne Paris Cit\'e, 5 place Jules Janssen, 92195 Meudon, France
\and Max Planck Institute for extraterrestrial Physics,
Giessenbachstraße~1, 85748 Garching, Germany
\and Max Planck Institute for Astronomy, K\"onigstuhl 17,
69117 Heidelberg, Germany
\and $1^{\rm st}$ Institute of Physics, University of Cologne,
Z\"ulpicher Straße 77, 50937 Cologne, Germany
\and Univ. Grenoble Alpes, CNRS, IPAG, 38000 Grenoble, France
\and Universidade de Lisboa - Faculdade de Ci\^encias, Campo Grande,
1749-016 Lisboa, Portugal
\and Faculdade de Engenharia, Universidade do Porto, rua Dr. Roberto
Frias, 4200-465 Porto, Portugal
\and European Southern Observatory, Karl-Schwarzschild-Straße 2, 85748
Garching, Germany
\and European Southern Observatory, Casilla 19001, Santiago 19, Chile
\and Observatoire de Gen\`eve, Universit\'e de Genève, 51 Ch. des
Maillettes, 1290 Versoix, Switzerland
\and Max Planck Institute for Radio Astronomy, Auf dem H\"ugel 69, 53121
Bonn, Germany
\and Sterrewacht Leiden, Leiden University, Postbus 9513, 2300 RA
Leiden, The Netherlands
\and Departments of Physics and Astronomy, Le Conte Hall, University
of California, Berkeley, CA 94720, USA
\and CENTRA - Centro de Astrof\'{\i}sica e
Gravita\c c\~ao, IST, Universidade de Lisboa, 1049-001 Lisboa,
Portugal
\and Dublin Institute for Advanced Studies,
31 Fitzwilliam Place, \mbox{Dublin 2}, Ireland
\and Department of Astronomy, University of Michigan, Ann Arbor, MI 48109, USA
\and European Space Agency, Space Telescope Science Institute,
3700 San Martin Drive, Baltimore MD 21218, USA
\and
Space Telescope Science Institute, Baltimore, MD 21218, USA
\and
STAR Institute, Universit\'e de Li\`ege, All\'ee du Six Ao\^ut 19c, B-4000 Li\`ege, Belgium
\and Department of Astronomy, California Institute of Technology, Pasadena, CA 91125, USA
\and 
Instituto de Astronomía, Universidad Nacional Autónoma de México, Apdo. Postal 70264, Ciudad de México 04510, Mexico
\and
School of Physics, Astrophysics Group, University of Exeter, Stocker Road, Exeter EX4 4QL UK
             }

   \date{Received September 15, 1996; accepted March 16, 1997}


  \abstract
    {}
  {To date, infrared interferometry at best achieved contrast ratios of a few times $10^{-4}$ on bright targets. GRAVITY, with its dual-field mode, is now capable of high contrast observations, enabling the direct observation of exoplanets. We demonstrate the technique on HR\ 8799, a young planetary system composed of four known giant exoplanets.}
   {
   We used the GRAVITY fringe tracker to lock the fringes on the central star, and integrated off-axis on the HR\ 8799\,e planet situated at 390\,mas from the star. Data reduction included post-processing to remove the flux leaking from the central star and to extract the coherent flux of the planet.    
   The inferred K band spectrum of the planet has a spectral resolution of 500. We also derive the astrometric position of the planet relative to the star with a precision on the order of 100$\,\mu$as.
   }
   {The GRAVITY astrometric measurement disfavors perfectly coplanar stable orbital solutions. A small adjustment of a few degrees to the orbital inclination of HR\ 8799\,e can resolve the tension, implying that the orbits are close to, but not strictly coplanar. The spectrum, with a signal-to-noise ratio of $\approx 5$ per spectral channel, is compatible with a late-type L brown dwarf. Using  Exo-REM synthetic spectra, we derive a temperature of $1150\pm50$\,K and a surface gravity of $10^{4.3\pm0.3}\,$cm/s$^{2}$. This corresponds to a radius of $1.17^{+0.13}_{-0.11}\,R_{\rm Jup}$ and a mass of $10^{+7}_{-4}\,M_{\rm Jup}$, which is an independent confirmation of mass estimates from evolutionary models. Our results demonstrate the power of interferometry for the direct detection and spectroscopic study of exoplanets at close angular separations from their stars.}
   {}


   \maketitle
%

\section{Introduction}

   \begin{figure*}
   \centering
   \includegraphics[width=0.98\textwidth]{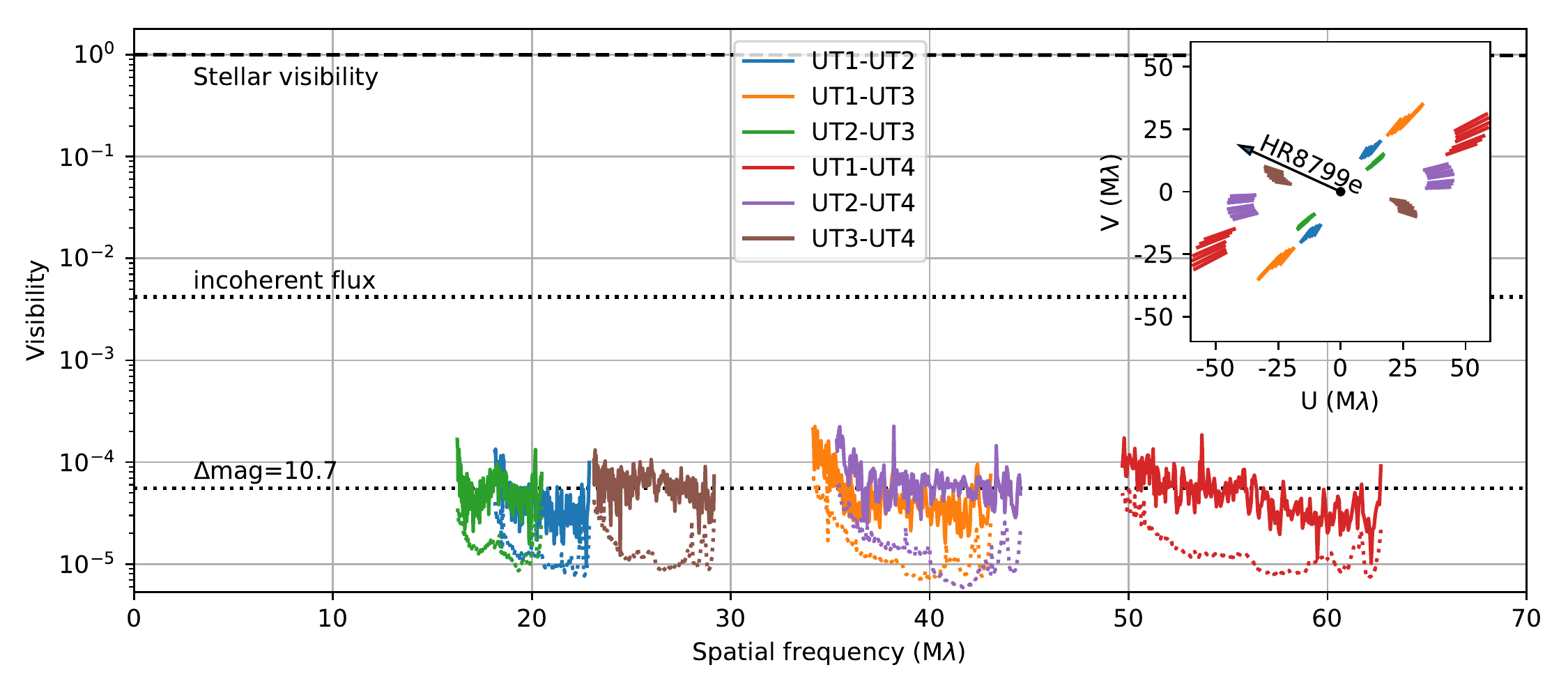}
   \caption{
    Colored lines are HR\ 8799\,e visibilities $|V_{\rm planet}|,$ as defined by Eq.~(\ref{eq:Vplanet}),  obtained from the ratio between the coherent flux observed on the planet and on the star. The underlying dotted colored lines correspond to the errors estimated by the pipeline.
   The theoretical stellar visibility corresponds to the black dashed line (uniform disk of diameter 0.342\,mas). The upper horizontal dotted black line corresponds to the observed incoherent flux (stellar flux leaking in the planet's field).
    The lower horizontal dotted black line corresponds to the theoretical visibility of a planet 10.7 magnitudes fainter than the star.  
    \textit{(Right inset)} Coverage of the spatial frequency plane,
   east is to the right. The arrow indicates the direction of the planet  situated to the northwest  of the star. 
     }
              \label{fig:UV}%
    \end{figure*}

 Obtaining accurate orbits, masses, and atmospheric spectra of directly imaged planets is
 key to determining their natures and, ultimately, their formation
 histories. Here we demonstrate the power of a new technique,
 using  optical interferometry, to obtain this information for an exoplanet as close
 as 390\,mas to its parent star.
 
Because they are better known, the spectra of brown dwarfs (BD) are often used as references to classify the young exoplanet atmospheres.
More precisely, the L-T transition is an important observable for  understanding the evolution of atmospheres as a function of temperature. At lower temperatures (<1200\,K), opacity changes due to the transition of CO to methane in chemical equilibrium, and the likely disappearance of silicate and iron clouds under the photosphere, makes the spectral appearance of T-type BDs bluer. 
On the contrary, young giant exoplanets of temperature $\approx 1000\,$K still have redder near-infrared colors typical of late-L BDs. This is explained by the relatively low surface gravity ($g\leq10^4$ cm/s$^2$), and hence larger scale heights, in planetary atmospheres. However, the exact cause (cloud properties and/or vertical chemical mixing) is not properly understood
\citep{allardAtmospheresVeryLowMass2012}. 
Once the problem of cloud formation and chemical processes are solved \citep{hellingDiskEvolutionElement2014,mosesCompositionYoungDirectly2016},
the determination of the molecular composition of exoplanet atmospheres will become a crucial tool toward understanding the formation process of planets: the atomic ratio or even isotope ratio will change depending on the conditions of formation. For example, the C-to-O ratio in the gas of a protoplanetary disk is predicted to increase outwards past the H$_2$O ($\approx 140$\,K), CO$_2$ ($\approx 50$\,K), or even CO  ($\approx 20$\,K) ice-lines \citep{obergEffectsSnowlinesPlanetary2011}:
 C/O  in the planet atmosphere should therefore change
depending on where in a disk planets form and how much gas and how many  planetesimals they accrete \citep{mordasiniIMPRINTEXOPLANETFORMATION2016}. Similarly, the D-to-H ratio in planets can be linked to the accretion of icy bodies \citep{feuchtgruberRatioAtmospheresUranus2013} and can be seen in molecular absorption spectra \citep{molliereDetectingIsotopologuesExoplanet2018}.

The study of cloud properties and composition requires spectral information, which can currently only be obtained by two means: transit spectroscopy or thermal emission spectroscopy. Transit spectroscopy is best for characterizing   planets in  close orbit around the host star, with  puffy (inflated) irradiated atmospheres \citep[][and references herein]{crossfieldObservationsExoplanetAtmospheres2015}.
 Thermal emission spectroscopy is more adapted to  young, self-luminous planets in orbits with a semimajor axis of a few tens of AU around the host star. Young planets are warm as they still possess excess entropy tracing back to the formation process \citep[e.g.,][]{marleauConstrainingInitialEntropy2014}. The difficulty with emission spectroscopy is that a planet's signal is contaminated by stellar photons, which vary in time with changing observing conditions, resulting in a spatially and spectrally varying speckle pattern.
One of the solutions for emission spectroscopy is to go to space to benefit from a stable point spread function. Another  is to use high contrast and high angular resolution observations on 8\,m to 10\,m class telescopes from the ground, and to deconvolve the image to remove the speckles by using spectro-spatial correlations. This is typically done using integral field spectroscopy and techniques like spectral differential imaging \citep{rameauDetectionLimitsSpectral2015} or angular differential imaging \citep{maroisAngularDifferentialImaging2006}.

With the technique presented in this paper, we go one step further by using the resolving power of the $\approx 100$-meter  baselines offered by optical interferometry to distinguish between the coherent flux originating from the star and from the planet. In Section~\ref{sec:log} we present the GRAVITY observation of HR\ 8799\,e and the data reduction. HR\ 8799 is a bright (K$=5.24$\,mag), nearby ($d=39.4\pm0.1$\,pc) A5 star. We know that at least four planets are orbiting the star \citep[the first three discovered by][]{maroisDirectImagingMultiple2008}.
In Section~\ref{sec:astr} we present a new astrometric measurement of the fourth planet (K $= 15.9$\,mag), discovered in 2010 by
 \citet{maroisImagesFourthPlanet2010}
at $368\pm9\,$mas from its host star.
The youth of the star  \citep[$\approx 30\,$Myr, member of the
Columba Association][]{maloBAYESIANANALYSISIDENTIFY2013}, implies that the planet is still warm from its initial gravitational energy. In Section \ref{sec:res}  we use the K-band spectra and photometry to constrain its spectral type, temperature, and radius.
Finally, in Section~\ref{sec:conclusion}, we summarize our results and briefly address the prospects of the interferometric technique.

\section{Observations and data reduction}
\label{sec:log}

   \begin{table}
   \scriptsize
      \caption[]{Observing log (data taken on 2018-Aug-28)}
\label{table:1}      
\centering                          
\begin{tabular}{l c r c c c c r}        
\hline\hline                 
            Target  & UT Time & DIT & NDIT & Seeing & $\tau_0$ & Airmass & par. angle\\
\hline                        
Planet  &      04:33:59  &  100\,s  &  10  &  0.8''  &  4.9\,ms & 1.48 & -164.3$^\circ$ \\
Star  &    04:51:27  &  1\,s  &  50  &  0.5''  &  7.1\,ms & 1.45 & -169.6$^\circ$ \\
Sky  &    04:52:56  &  1\,s  &  50  &  0.5''  &  6.7\,ms & 1.45 & -170.0$^\circ$ \\
Planet  &    04:54:35  &  100\,s  &  10  &  0.6''  &  6.2\,ms & 1.45 & -170.6$^\circ$ \\
Planet  &    05:11:55  &  100\,s  &  10  &  0.6''  &  6.2\,ms & 1.44 & -176.0$^\circ$ \\
Sky &    05:29:27  &  100\,s  &  10  &  0.4''  &  8.5\,ms & 1.44 & 178.5$^\circ$ \\
Planet  &    05:46:56  &  100\,s  &  10  &  0.5''  &  6.3\,ms & 1.44 & 172.9$^\circ$ \\
Planet  &    06:07:13  &  100\,s  &  10  &  0.6''  &  5.6\,ms & 1.47 & 166.7$^\circ$ \\
Star  &    06:24:42  &  1\,s  &  50  &  0.8''  &  4.2\,ms & 1.50 & 161.5$^\circ$ \\
Sky &    06:26:10  &  1\,s  &  50  &  0.8''  &  3.9\,ms & 1.50 & 161.1$^\circ$ \\
Planet  &    06:28:04  &  100\,s  &  10  &  0.7''  &  4.9\,ms & 1.50 & 160.5$^\circ$ \\
Planet  &    06:45:25  &  100\,s  &  10  &  0.7''  &  6.2\,ms & 1.55 & 155.7$^\circ$ \\
Sky  &    07:04:02  &  100\,s  &  10  &  1.0''  &  3.6\,ms & 1.62 & 150.9$^\circ$ \\
\hline                                   
\end{tabular}
Notes: $\tau_0$ is the coherence time in the visual (500\,nm); par.\ angle is the parallactic angle.
\end{table}

The observations were obtained with the VLTI using the four 8-meter Unit Telescopes
and the GRAVITY instrument  \citep{eisenhauerGRAVITYObservingUniverse2011,gravitycollaborationFirstLightGRAVITY2017} on 28  August 2018.
GRAVITY can observe two objects located in the VLTI field of view by simultaneously injecting, at each telescope coudé focus, the light of each object 
 into a separate single-mode fiber. The two fibers have an effective field of view equal to $\approx 60\,$mas, the K-band diffraction limit of a single telescope \citep{pfuhlFiberCouplerBeam2014}.
 Each object is thus interferometrically observed separately, but an ultraprecise laser telemetry constantly monitors the differential optical path between the two objects. 
The first fiber of GRAVITY was placed on the star for fringe tracking \citep{lacourGRAVITYFringeTracker2019} and phase referencing. The second fiber was centered sequentially on the planet  and on the star,  situated at $\approx 390\,$mas from each other. This second fiber fed the science spectrometer  configured at  medium resolution (R$=500$). Observations of the star were used to calibrate the observation of the planet.
Because of the faintness of the planet, a detector integration time (DIT) of 100\,s was required. As the star is more than 10 mag brighter than the planet in K band, a DIT of 1\,s was sufficient for the science channel observations of the star. Seeing conditions were average to good. The log of the observations is presented in Table~\ref{table:1}, where exposures on the planet are shown along the sky and stellar calibration exposures.

The frequency plane and the amplitude of the planet's visibilities are presented in Fig.~\ref{fig:UV}. The colored dotted lines below the visibilities are the errors estimated by the pipeline. The mean S/N per spectral channel is $\approx 5$.
The detailed data reduction procedure will be presented by Nowak et al. (in preparation). The main steps are as follows:
\begin{enumerate}
\item Extraction of the coherent flux (the  VISDATA) for individual files using the ESO GRAVITY pipeline\footnote{The pipeline in its version 1.1.2 is currently available at \url{https://www.eso.org/sci/software/pipelines/gravity}.};
\item  Derivation of the position of the planet with respect to the star by fitting the coherent flux with  models of the coherent flux from the star and from the planet;
\item  Removal of the coherent flux of  stellar origin by linear decomposition on the  models of step 2. This step assumes the position of the planet from step 2; 
\item  Normalization in phase and amplitude  of the remaining coherent flux by the coherent flux observed on the star multiplied by the theoretical visibility function of the star. This gives the complex visibility of the planet
\begin{equation}
V_{\rm planet}=\frac{{\rm VISDATA}_{\rm planet}}{{\rm VISDATA}_{\rm star}^*} \times  \frac{2 J_1(\pi \theta_{\rm star} u  )}{\pi \theta_{\rm star} u  }
\label{eq:Vplanet}
,\end{equation}
where $J_1$ is a Bessel function of the first kind, of order 1; $\theta_{\rm star}$ is the stellar diameter; and $u$ is the spatial frequency in rad$^{-1}$. The amplitude of 
 $V_{\rm planet}$ is plotted as solid curves in Fig.~\ref{fig:UV};
\item Retrieval of the spectrum of the planet by assuming a diameter for the planet and a synthetic stellar spectrum:
\begin{equation}
F_{\rm planet} = \frac{ |V_{\rm planet}| F_{\rm star}  }{2 J_1(\pi \theta_{\rm planet} u  )/\pi \theta_{\rm planet} u }
\label{eq:Fspectra}
.\end{equation}
\end{enumerate}
The stellar diameter is assumed to be $\theta_{\rm star} = 0.342\pm0.008\,$mas  \citep{bainesCHARAArrayAngular2012} and is plotted
 as the black dashed line in Fig.~\ref{fig:UV}.  The planet diameter is assumed to be negligible at the resolution of the interferometer. For the star, we used a BT-NextGen model ($T=7400\,$K, [Fe/H]=-0.5, and log$(g)=4.0$) from \citet{hauschildtNEXTGENModelAtmosphere1999}, scaled for a K-band flux of $3.191\times10^{-12} $\,Wm$^{-2}\mu $m$^{-1}$.
 
 The upper dotted black line corresponds to the average residual flux from the star entering the science spectrometer. The lower dotted black line corresponds to the theoretical flux of an unresolved source with a 10.7 difference in magnitude.

%

\section{Relative astrometry}
\label{sec:astr}

   \begin{table}
   \small
      \caption[]{Astrometry on HR\ 8799\,e}
\label{table:2}      
\centering                          
\begin{tabular}{l c c c}        
\hline\hline                 
             MJD & $\Delta$RA & $\Delta$Dec & Covariance \\
             & (mas) & (mas)  & (mas$^2$)\\
\hline                        
    58358.190  & -357.63 & 163.59 \\
    58358.205  & -357.68 & 163.63 \\
    58358.217 &  -357.54 & 163.05 \\
    58358.241 &  -357.58 & 163.28 \\
    58358.255 &   -357.61 & 163.12 \\
    58358.269  &  -357.62 & 163.22 \\
    58358.282 &  -357.80 & 163.41 \\
\hline                                   
    Global     & $-357.64\pm0.07$   & $163.34\pm0.18$ & -0.00668 \\
\hline                                   
\end{tabular}
\end{table}

   \begin{figure}
   \centering
   \includegraphics[width=0.48\textwidth]{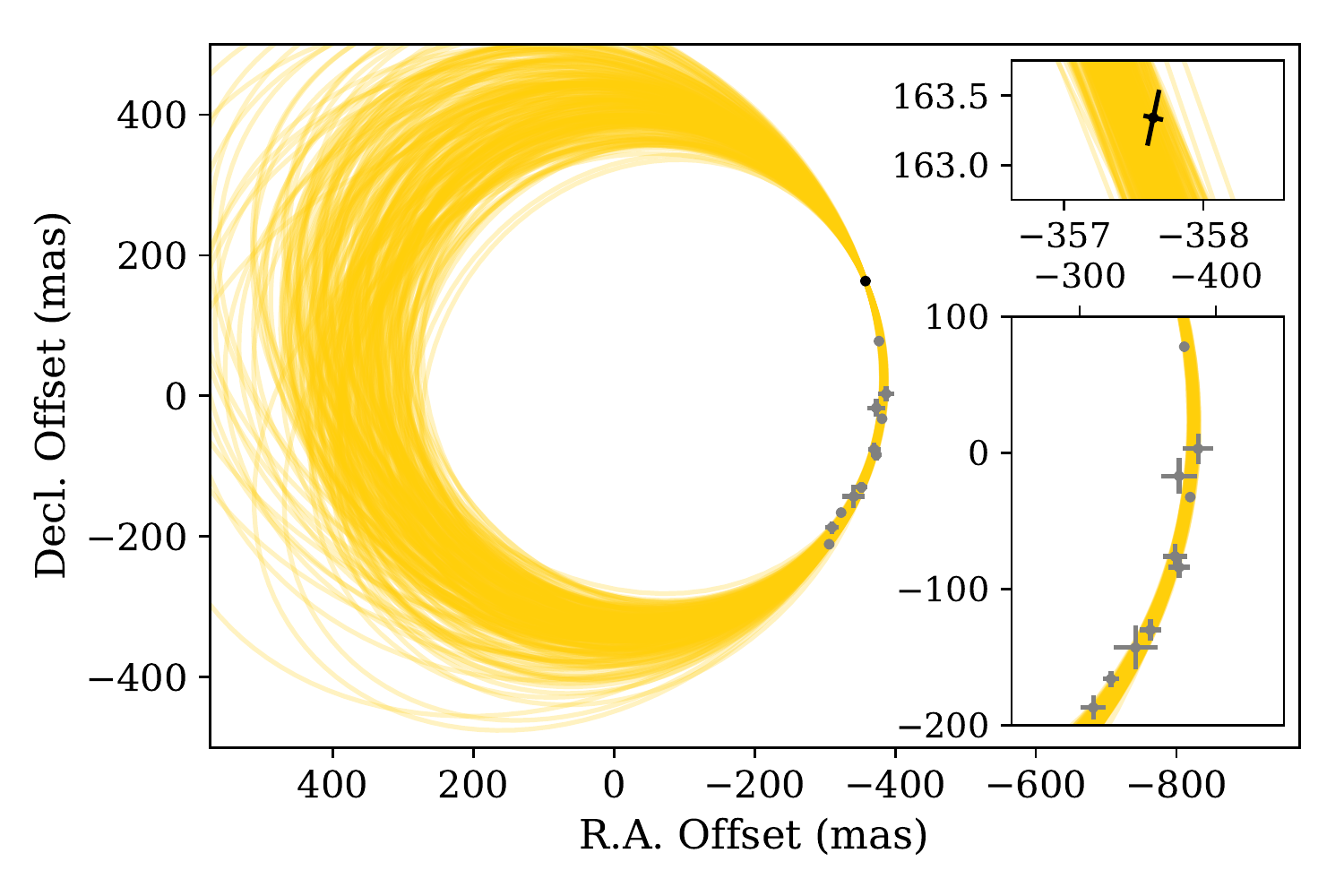}
   \caption{
   Keplerian orbit fit of HR\ 8799\,e. The black point is the GRAVITY measurement, and the gray points are from previous astrometry  \citep{konopackyASTROMETRICMONITORINGHR2016,wangDynamicalConstraintsHR2018}. The yellow lines are 250 random draws from the posterior. The orbit determination is currently limited by the mas-level astrometry of the previous epochs \textit{(Top inset)}. Image magnified by $\sim$200x to display the uncertainties in the GRAVITY astrometry. The plotted error bars are rotated to be aligned with the principal axes of the error ellipse. \textit{(Bottom inset)} Image magnified  by $\sim$2x to display the uncertainties in the previous measurements.
   }
              \label{fig:position}%
    \end{figure}

   \begin{figure*}
   \centering
   \includegraphics[width=0.95\textwidth]{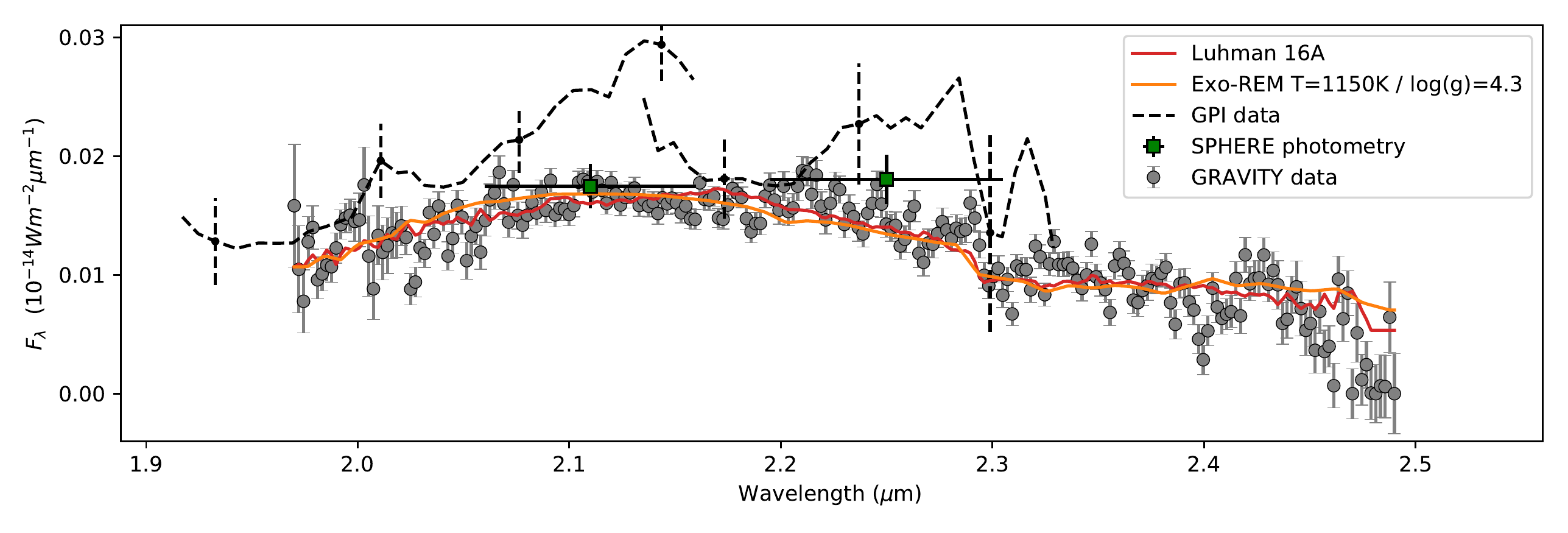}
   \caption{GRAVITY K band spectrum of HR8799\,e at spectral resolution 500 (gray points). The red curve is  the X-SHOOTER spectrum of the brown dwarf Luhman\,16\,A from \citet{lodieuVLTXShooterSpectroscopy2015b}, smoothed to the GRAVITY resolution. The reduced $\chi^2_{\rm red}$ is 2.4 (over 236 degrees of freedom). The orange curve is the best fit of the Exo-REM models from
   \citet{charnaySelfconsistentCloudModel2018}. The reduced $\chi^2_{\rm red}$ is 2.7.
 The dashed curve is the K-band GPI spectrum from \citet{greenbaumGPISpectraHR2018}.
The square dots are the SPHERE photometry from  \citet{zurloFirstLightVLT2016}.
  }
              \label{fig:spectra}%
    \end{figure*}

In the same way as we can disentangle the complex coherent energy from the star and the planet, it is also possible to fit the wavelength dependence of the phase, which is tantamount to measuring a separation. Each fit, for each baseline, gives a $\chi^2$ minimum for an optimal optical path difference (OPD). This OPD corresponds to an angular separation projected in the direction of the baseline vector.
Several of these optimal OPDs are necessary to derive a position.
By combining all the baselines together, we can use  each exposure file separately, giving the seven optimal positions listed in Table~\ref{table:2}. The global minimum is at $\Delta {\rm RA}= -357.64\pm0.07\,$mas  and $\Delta {\rm Dec}=163.34\pm0.18\,$mas with highly elliptical uncertainty (covariance of -0.00668\,mas$^2$). Along the longest baseline (position angle, PA=78 degrees) the 1$\sigma$ uncertainty is 55$\,\mu$as. Orthogonal to that baseline (PA= 168 degrees) the uncertainty is 190$\,\mu$as. The plate scale and true north error is negligible at that level ($>50\,\mu$as) as the spatial frequencies are defined by the physical position of the telescopes. Atmospheric dispersion is also negligible. A detailed description of the error terms of interferometric astrometry is presented in  \citet{lacourReachingMicroarcsecondAstrometry2014}. 

As this astrometry is an order of magnitude more precise  than the best measurements made by direct imaging instruments \citep{wangDynamicalConstraintsHR2018}, we investigate the orbital constraints provided by this datapoint. We fit a single Keplerian orbit by combining this measurement with the astrometry reported in \citet{konopackyASTROMETRICMONITORINGHR2016} and \citet{wangDynamicalConstraintsHR2018}. Given the assymetry in the GRAVITY measurement, we fit for the location of the planet at the GRAVITY epoch in a rotated frame that is aligned with the two principal axes of the error ellipse (top inset of Fig. \ref{fig:position}). We use the parallel-tempered Markov chain Monte Carlo sampler \citep{foreman-mackeyEmceeMCMCHammer2013,vousdenDynamicTemperatureSelection2016} in the orbit fitting code \texttt{orbitize} \citep{sarahbluntSbluntOrbitizeExpand2019}
to estimate the orbital parameters and find a semimajor axis of $16.4^{+2.1}_{-1.1}$~AU, an eccentricity of $0.15 \pm 0.08$, and an inclination of $25^{\circ} \pm 8^{\circ}$. A single 100$\,\mu$as precision point is able to significantly constrain the position of the planet at the epoch of observation, but the determination of the planet's velocity, acceleration, and orbital properties are still dominated by the mas-level uncertainties in the previous astrometry. We therefore defer a thorough dynamical study to a time when multi-epoch orbital monitoring of the planet with VLTI/GRAVITY has been obtained.

We can also compare the location of the planet measured by GRAVITY with the $\Delta{\rm RA}= -352.6^{+3.1}_{-2.6}\,$mas and $\Delta{\rm Dec}= -157.9 \pm 1.8\,$mas  predicted by the dynamically stable coplanar solutions from \citet{wangDynamicalConstraintsHR2018}. The positions are inconsistent by 5~mas in both axes and none of the 9792 stable coplanar orbits are consistent with our measurement at the 3$\sigma$ level. With this single astrometric point, we are therefore able to disfavor dynamically stable configurations in which the four planets are perfectly coplanar.
Changing the inclination of HR\ 8799\,e by $\approx -2^{\circ}$ accounts for this 5~mas difference, and \citet{wangDynamicalConstraintsHR2018} did find 14 stable non-coplanar orbits with mutual inclinations of less $8^{\circ}$ out of 20 million trials. We note that given the uncertainties in the orbital planes of the other three planets, we cannot pinpoint the mutual inclinations of the planets in this simple analysis. Continued monitoring of the orbit with GRAVITY can further constrain the planet's orbital elements, allowing a search for dynamically stable non-coplanar orbital solutions to be computationally tractable and providing more accurate constraints on the masses of the multiple planets.

\section{Atmosphere of HR\ 8799\,e}
\label{sec:res}


   \begin{figure}
   \centering
   \includegraphics[width=0.49\textwidth]{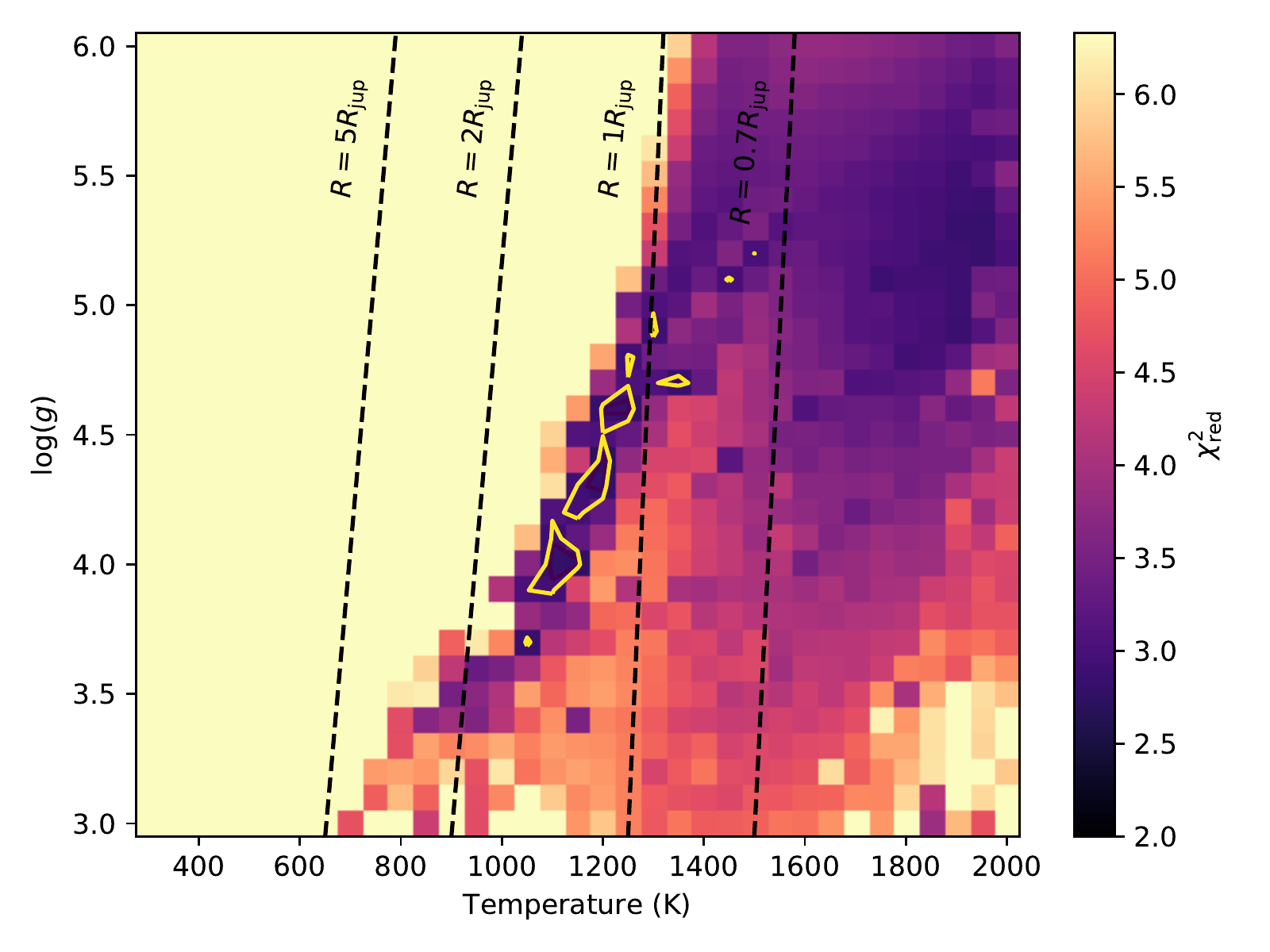}
   \caption{ Reduced $\chi^2$ as a function of effective temperature and surface gravity from a grid of Exo-REM models \citep{charnaySelfconsistentCloudModel2018}. The yellow contours correspond to the $5\sigma$ error, indicating a valley of possible temperatures between 1100 and 1200\,K. The vertical lines correspond to the planet's radius from the model K-band luminosity assuming a distance of 39.4\,pc.
  }
              \label{fig:BT}%
    \end{figure}


The GRAVITY spectrum of HR\ 8799\,e was obtained by multiplying the visibility of the planet with the theoretical spectrum of the star. This is following  Eq.~(\ref{eq:Fspectra}) and assuming $\theta_{\rm planet}=0$.
The resulting spectrum is represented as the gray points in Fig.~\ref{fig:spectra}. The CO-band head at 2.29\,$\mu$m is the most prominent feature. As already mentioned by \citet{konopackyDetectionCarbonMonoxide2013} and \citet{wangDetectingWaterAtmosphere2018} for HR8799\,c, no clear CH$_4$ absorption is seen, in agreement with a typical L-type BD spectrum. Using H and K band GPI spectra, \citet{greenbaumGPISpectraHR2018}  obtained a best fit with the spectrum of the brown dwarf WISE J1049-5319A
\citep[also called Luhman\,16\,A from ][]{luhmanDiscoveryBinaryBrown2013} of spectral type L7.5. The fit is equally good with the GRAVITY spectrum, and gives a reduced $\chi_{\rm red}^2$ of 2.4.

We fitted the catalog of BD spectra from the Montreal library \citep{gagneBANYANVIINew2015,robertBrownDwarfCensus2016} to try to narrow down the spectral type from K-band spectroscopy only. With a reduced $\chi^2$ of 2.4, the best fit indicates a spectral type close to L7 BD, in agreement with  \citet{bonnefoyFirstLightVLT2016} and  \citet{greenbaumGPISpectraHR2018}. A T-type BD spectrum is clearly ruled out. The reduced $\chi^2$ increases to 3 for spectral types $\approx$L4, which is significant with 230 degrees of freedom. Similarly, we fitted a grid of BT-Settl 2014 synthetic spectra \citep{baraffeNewEvolutionaryModels2015} to derive a temperature and a surface density. The best fit was obtained for a temperature of 1400\,K and a surface gravity of $10^4$\,cm/s$^2$. This corresponds to a planetary radius of $0.8\,R_{\rm Jup}$.

This radius being incompatible with evolutionary models, we turned to the Exo-REM model \citep{baudinoInterpretingPhotometrySpectroscopy2015,charnaySelfconsistentCloudModel2018}. We found that values of $1150\pm50\,$K and log(g)$=4.3\pm0.3$ (error bars 3$\sigma$) correctly reflect the spectrum in the K band (Fig.~\ref{fig:BT}). According to the luminosity estimated by the model, it corresponds to a radius of $R=1.17^{+0.13}_{-0.11}\,R_{\rm Jup}$. This gives a model-dependent estimate of the mass of HR\ 8799\,e of $10^{+7}_{-4}\,M_{\rm Jup}$.
Simulations with Exo-REM predict that the LT transition occurs at a lower effective temperature for exoplanets than for field brown dwarfs, due to effects of pressure on the formation of iron and silicate clouds. This trend is apparent in Fig.\ \ref{fig:BT}, where the LT transition corresponds to a sudden increase in $\chi^2$ and occurs at an effective temperature just 100 K lower than our best fit.

\section{Summary and conclusions}
\label{sec:conclusion}

Interferometric astrometry, an order of magnitude more accurate than direct imaging, opens new possibilities to study the dynamics of planetary systems. With just a single data point from GRAVITY, we can strongly disfavor perfectly coplanar stable orbits for the HR 8799 planets. As the dynamics probe the masses, formation history, and the future system architecture, interferometric orbital monitoring at the 10-100$\,\mu$as level can significantly improve our understanding of directly imaged systems.

Based on the K-band spectrum, we confirm a
spectral type ($\approx$L7), equivalent to a higher temperature BD.
The discrepancy between spectral type \citep[T$ >1400\,K$;][]{schweitzerEffectiveTemperaturesLate2002} and effective temperature derived from wide-band photometry (T$ <1200\,K$) can be solved by using models taking  the lower surface gravity into account.
 It is interesting to note that the GRAVITY K-band spectrum does constrain this
  low surface gravity, as shown by the residual map in Fig.~\ref{fig:BT}. We determine a surface gravity compatible with a $10\,M_{\rm Jup}$ planet.

The interferometric technique brings unique possibilities to characterize exoplanets. With the technique described here, any planet with $Kmag \lessapprox 19$, $\Delta Kmag \lessapprox 11$, and separation $\gtrapprox 100$\,mas is, in theory, observable with GRAVITY. The numbers are still to be refined, but it would mean that GRAVITY could observe most of the known imaged planets, and maybe in the near future planets detected by radial velocity.
Futhermore, the  good normalization of the continuum  spectrum offers new ways to measure the column density of molecules without the need for smoothing and cross-correlation \citep[e.g.,][]{snellenFastSpinYoung2014,konopackyDetectionCarbonMonoxide2013}.  
Finally, the idea that an interferometer can resolve the surface of exoplanets,  giving radius and resolving clouds patchiness, is now becoming more plausible. However, it would require an interferometer with baselines on the order of 10\,km. This could be a goal for ESO after ELT construction.


\begin{acknowledgements}
GRAVITY was developed via a
collaboration of the Max Planck Institute for Extraterrestrial Physics,
LESIA of Paris Observatory and IPAG of Université Grenoble Alpes / CNRS,
the Max Planck Institute for Astronomy, the University of Cologne, the
Centro Multidisciplinar de Astrofisica Lisbon and Porto, and the European
Southern Observatory.
Part of this work was supported by the European Union under ERC grant 639248 LITHIUM. J.W. is supported by the Heising-Simons Foundation 51~Pegasi~b postdoctoral fellowship. A.A., P.G., and N.A. acknowledge funding from Fundação para a Ciência e Tecnologia through grants  PTDC/CTE-AST/116561/2010, SFRH/BD/52066/2012, COMPETE
FCOMP-01-0124-FEDER-019965,  UID/FIS/00099/2013, and SFRH/BSAB/142940/2018.
\end{acknowledgements}

%
   \bibliographystyle{aa} 
   \bibliography{MyLibrary} 

%
%
%

%
%
%
%

\end{document}